\documentclass{Interspeech2024}

\usepackage{mathtools}
\usepackage{amsmath,graphicx}
\usepackage[english]{babel}
\usepackage[table, dvipsnames]{xcolor}

\interspeechcameraready

\title{Cross-modal Knowledge Transfer Learning as Graph Matching Based on Optimal Transport for ASR}

\address{
	\vspace{-6mm}
	\textit{Xugang Lu$^{1*}$, Peng Shen$^{1}$, Yu Tsao$^{2}$, Hisashi Kawai$^1$} \\
	\thanks{*Acknowledgment: The work is partially supported by JSPS KAKENHI No. 24K15004.}\\
	\vspace{-4mm}
	1. National Institute of Information and Communications Technology, Japan\\
	2. Research Center for Information Technology Innovation, Academia Sinica, Taiwan}
\email{xugang.lu@nict.go.jp}


\usepackage{comment}

\begin{document}

\maketitle

\vspace{-8mm}	
\begin{abstract}

Transferring linguistic knowledge from a pretrained language model (PLM) to acoustic feature learning has proven effective in enhancing end-to-end automatic speech recognition (E2E-ASR). However, aligning representations between linguistic and acoustic modalities remains a challenge due to inherent modality gaps. Optimal transport (OT) has shown promise in mitigating these gaps by minimizing the Wasserstein distance (WD) between linguistic and acoustic feature distributions. However, previous OT-based methods overlook structural relationships, treating feature vectors as unordered sets. To address this, we propose Graph Matching Optimal Transport (GM-OT), which models linguistic and acoustic sequences as structured graphs. Nodes represent feature embeddings, while edges capture temporal and sequential relationships. GM-OT minimizes both WD (between nodes) and Gromov-Wasserstein distance (GWD) (between edges), leading to a fused Gromov-Wasserstein distance (FGWD) formulation. This enables structured alignment and more efficient knowledge transfer compared to existing OT-based approaches. Theoretical analysis further shows that prior OT-based method in linguistic knowledge transfer can be viewed as a special case within our GM-OT framework. We evaluate GM-OT on Mandarin ASR using a CTC-based E2E-ASR system with a PLM for knowledge transfer. Experimental results demonstrate significant performance gains over state-of-the-art models, validating the effectiveness of our approach.
\end{abstract}

\section{Introduction}
The pretrained language models (PLMs) learned from large text corpora encodes rich linguistic knowledge, and has greatly advanced the performance of automatic speech recognition (ASR)  with end-to-end (E2E) acoustic models \cite{Chan2016,Kim2017,Hori2017,Watanabe2017,RNNTASR}. While enhancing ASR in decoding, typically as external language models (LMs) for tasks like beam search and rescoring \cite{BERTScore,MLMScore}, external LMs introduce computational inefficiencies and hinder parallel decoding. A key challenge is transferring linguistic knowledge into the acoustic encoding stage, enabling ASR without external LM dependencies. This work explores knowledge transfer from PLMs to a temporal connectionist temporal classification (CTC)-based ASR \cite{CTCASR}. While some E2E-ASR methods integrate linguistic information \cite{HierarchicalCTC,intermediateCTC}, leveraging PLMs like BERT \cite{BERT} has shown promise \cite{FNAR-BERT,NARBERT,KuboICASSP2022,Choi2022,Futami2022,Higuchi2023,CIFBERT1,CIFBERT2, wav2vecBERTSLT2022,CTCBERT1,CTCBERT2}. In their models, cross-attention based alignment model in transformer-based decoding \cite{Transformer}, or Continuous Integrate-and-Fire (CIF) neural network-based alignment model \cite{CIFBERT1, CIFBERT2} are applied in integrating linguistic knowledge with acoustic models. However, due to modality gaps, aligning acoustic and linguistic feature representations still remains a challenge. 

Optimal transport (OT) provides a natural framework for cross-modal alignment by minimizing feature distributional discrepancies \cite{VillanoBook}. OT has been applied to domain adaptation \cite{CourtyNIPS2017}, spoken language recognition, speech enhancement \cite{LuICASSP2021,Lin2021,ICLR2023}, and speech understanding \cite{Cross2021,ACL2023,ICML2023}. Recent studies explore OT-based knowledge transfer learning for ASR \cite{ASRU2023Lu,SLT2024Lu}, but most treat feature representations as unordered sets, ignoring the structured nature of speech. In reality, acoustic features exhibit temporal coherence, and linguistic features follow a sequential order. We hypothesize that incorporating these structural properties into OT can improve alignment accuracy and efficiency in linguistic knowledge transfer. To address this, we introduce Graph Matching Optimal Transport (GM-OT), framing cross-modal alignment as a graph-matching problem. Each utterance is represented as two graphs, one in acoustic space and one in linguistic space with nodes as feature embeddings and edges capturing temporal and sequential structures. Effective alignment requires matching both nodes and edges, ensuring feature similarity and structural coherence, ultimately improving E2E-ASR performance.

Our main contributions are as follows: 1. We propose a novel GM-OT framework for linguistic knowledge transfer in E2E-ASR, leveraging graph-based optimal transport for structured feature alignment. 2. We provide theoretical insights, showing that existing OT-based alignment and transfer learning methods can be derived as special cases of our GM-OT framework. 3. We conduct experiments on a CTC-based E2E-ASR system incorporating a PLM for knowledge transfer. Our results demonstrate significant performance gains in linguistic knowledge transfer learning for E2E-ASR. 
\vspace{-2mm}
\section{Proposed method}
\label{sec:proposed}
\begin{figure}[tb]
	\centering
	\includegraphics[width=6cm, height=3.5cm]{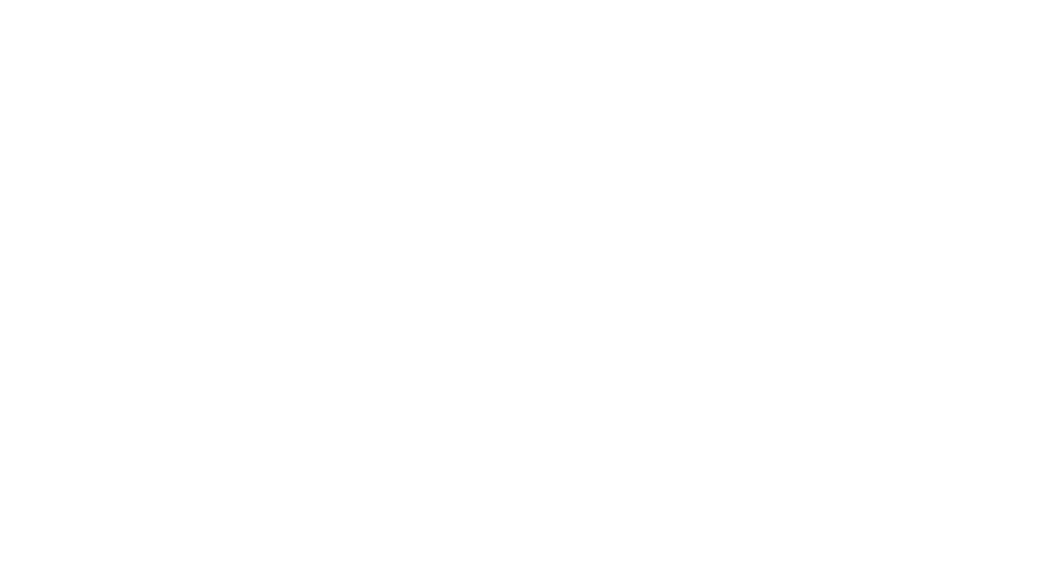}
	\caption{The proposed cross-modal alignment and knowledge transfer model for ASR based on GM-OT.}
	\vspace{-2mm}		
	\label{fig:fig1}
\end{figure}

The proposed model framework is illustrated in Fig. \ref{fig:fig1}. In this framework, two encoders, i.e., acoustic encoder and linguistic encoder are used for exploring acoustic and linguistic feature representations. Between the two encoders, there is one `Adapter' module to align acoustic feature dimension with the linguistic one (gray blocks in Fig. \ref{fig:fig1}). The main difference between our work and other studies \cite{ASRU2023Lu} is that our work introduces a GM-OT alignment module for matching these representations. Before detailing GM-OT, we briefly explain the feature exploration blocks. 
\subsection{Feature exploration in acoustic and linguistic encoders}
In acoustic encoder, convolution-based down-sampling is applied in order to reduce the duration lengths of acoustic sequence (a down-sampling rate of $4$ was used in this paper). After down-sampling, a positional encoding is applied in order to keep temporal position sensitivity in acoustic encoding. The representation from acoustic encoder is obtained as:
\begin{equation}			
		{\bf H}_{{\rm enc}}  = {\rm AcousticEncoder}\left( {{\bf X} } \right) \in \mathbb{R}^{l_a  \times d_a },  	
	\label{eq:conformer}
\end{equation}
where $l_a$ and $d_a$ are temporal length and dimension of the acoustic feature vectors, respectively. A linear projection termed `FC2' in the `Adapter' module is utilized to perform a feature dimension matching transform as:
\begin{equation}
	{\bf H}^{\rm A}  = {\rm FC}_{\rm 2} \left( {{\bf H}_{{\rm enc}} } \right) \in \mathbb{R}^{l_a  \times d_t }, 
	\label{eq:FC2}
\end{equation}
where $d_t$ corresponds to linguistic feature dimension. 

The context-dependent linguistic feature representation is explored from a linguistic encoder. The process is formulated as: 
\begin{equation}
	\begin{array}{l}
		{\bf Z}_0  = \left[ {{\rm CLS, }{\bf y}_{{\rm token}} ,{\rm SEP}} \right] \\ 
		{\bf Z}^{\rm L}  = {\rm LinguisticEncoder} \left( {{\bf Z}_0 } \right) \in \mathbb{R}^{l_t  \times d_t }, \\ 		
	\end{array}
	\label{eq:bert}
\end{equation}
where ${\bf y}_{{\rm token}}$ are word piece-based tokens, e.g., tokens by tokenization process of bidirectional encoder representation from transformers (BERT) \cite{BERT}. Token symbols `CLS' and `SEP' represent the start and end of an input token sequence. In Eq. (\ref{eq:bert}), $l_t$ denotes the sequence length, and $d_t$ represents feature dimension of text encoding representation. 

\subsection{Feature alignment based on Optimal Transport}

\begin{figure}[tb]
	\centering
	\includegraphics[width=6cm, height=3.5cm]{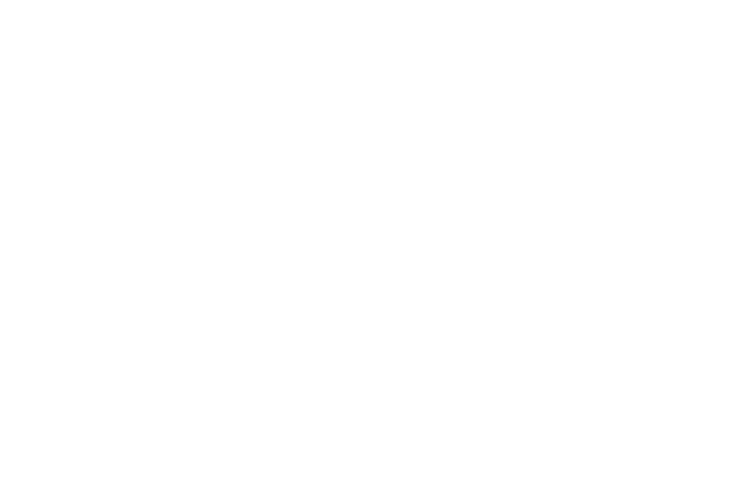}
	\caption{Feature alignment between acoustic and linguistic modalities, (a) OT on unordered set, (b) OT on set with pair-wised relevance, (c) OT on set with temporal consistent topology structure.}
	\vspace{-4mm}		
	\label{fig:fig2}
\end{figure}

After extracting acoustic and linguistic features, alignment is needed for efficient knowledge transfer. OT maps one probability distribution to another with minimal transport cost \cite{VillanoBook} and has been used for representation alignment in ASR transfer learning \cite{ASRU2023Lu}. However, their usage of OT treats feature vectors as an unordered set, ignoring speech and linguistic structures. We propose a new perspective, modeling each utterance as graphs in acoustic and linguistic spaces, where feature vectors are nodes and their relationships form edges. Fig. \ref{fig:fig2} illustrates the difference between vanilla OT and our graph-matching-based OT (GM-OT) for alignment.

In Fig. \ref{fig:fig2} (a), traditional OT uses Wasserstein distance (WD) on an unordered set of independent nodes. In Fig. \ref{fig:fig2} (b), OT considers pair-wised node matching (edges). Our proposed GM-OT in Fig. \ref{fig:fig2} (c) incorporates the set’s topology, aligning both nodes and edges in an ordered structure.
Before detailing Fig. \ref{fig:fig2}, we define the OT matching inputs: acoustic and linguistic features ${\bf H}^{\rm A}  = \left[ {{\bf h}_1 ,{\bf h}_2 ,...,{\bf h}_i ,...,{\bf h}_{l_a } } \right]$, ${\bf Z}^{\rm L}  = \left[ {{\bf z}_1 ,{\bf z}_2 ,...,{\bf z}_k ,...,{\bf z}_{l_t } } \right]$, 

where $l_a$ and $l_t$ are temporal sizes of the two sequences. Suppose the two sequences are sampled from two probability distributions with weight vectors ${\bf a}  = \left[ {a_1 ,a_2 ,...,a_i ,...,a_{l_a } } \right]$ and 
${\bf b}  = \left[ {b_1 ,b_2 ,...,b_k ,...,b_{l_t } } \right]$. ($a_i  = {1 \mathord{\left/
		{\vphantom {1 {l_a }}} \right.
		\kern-\nulldelimiterspace} {l_a }},b_k  = {1 \mathord{\left/
		{\vphantom {1 {l_t }}} \right.
		\kern-\nulldelimiterspace} {l_t }}$ as uniform distributions if no prior information is available). 	
\subsubsection{Matching on nodes: Minimizing Wasserstein distance}
The matching on un-ordered sets is explained in Fig. \ref{fig:fig2} (a). The Wasserstein distance (WD) (or OT distance) between the two sequences is defined on nodes as:       
\begin{equation}
L_{\rm WD} \mathop  = \limits^\Delta  \mathop {\min }\limits_{\gamma  \in \prod {\left( {{\bf H}^{\rm A} ,{\bf Z}^{\rm L} } \right)} } \left\langle {{{\bf D}^{{\rm AL}} },\gamma} \right\rangle, 
\label{eq:ot}
\end{equation}
where ${\gamma}$ is a transport coupling set defined as:
\begin{equation}	
\prod {\left( {{\bf H}^{\rm A} ,{\bf Z}^{\rm L} } \right)} \mathop  = \limits^\Delta  \left\{ {\gamma  \in R_ + ^{l_a  \times l_t } \left| {\gamma {\bf 1}_{l_t }  = {\bf a},\gamma ^T {\bf 1}_{l_a }  = {\bf b}} \right.} \right\}
\label{eq:coupling}
\end{equation}
In Eq. (\ref{eq:coupling}), ${\bf 1}_{l_a }$ and ${\bf 1}_{l_t }$ are vectors of ones with dimensions $l_a$ and $l_t$, respectively. In Eq. (\ref{eq:ot}), ${{\bf D}^{{\rm AL}} }$ is a distance matrix (or ground metric) with element ${d_{i,j}^{\rm AL} }$ defined as cross-modal pair-wised cosine distance:
\begin{equation}
	{\bf D}^{\rm AL}({\bf h}_i ,{\bf z}_j ) = d_{i,j}^{\rm AL} \mathop  = \limits^\Delta  1 - \cos ({\bf h}_i ,{\bf z}_j )
	\label{eq:Cost}
\end{equation}
A fast solution for OT has been introduced through the celebrated entropy-regularized OT (EOT) \cite{Cuturi2013} where the EOT loss is defined as:
\begin{equation}
	L_{\rm WD}^{\beta } \left( {{\bf H}^{\rm A} ,{\bf Z}^{\rm L} } \right)\mathop  = \limits^\Delta  \mathop {\min }\limits_{\gamma  \in \prod {\left( {{\bf H}^{\rm A} ,{\bf Z}^{\rm L} } \right)} } \left\langle {{\bf D}^{\rm AL},\gamma } \right\rangle  - \beta H\left( \gamma  \right), 
	\label{eq:EOTloss}  
\end{equation}
where $\beta $ is a regularization coefficient, and $H\left( \gamma  \right) =  - \sum\limits_{i,j} {\gamma _{i,j} \log \gamma _{i,j} } $ is entropy of coupling matrix. The solution of Eq. (\ref{eq:EOTloss}) can be implemented with Sinkhorn algorithm as in \cite{Cuturi2013}.

\vspace{-2mm}
\subsubsection{Matching on edges: Minimizing Gromov-Wasserstein distance}
For matching edges which considers the relevance of pair-wised nodes between modalities, it is illustrated in Fig. \ref{fig:fig2} (b). It is formulated as minimizing the Gromov-Wasserstein distance (GWD) \cite{Peyre2016}:
\vspace{-2mm}
\begin{equation}
	L_{{\rm GWD}} \mathop  = \limits^\Delta  \mathop {\min }\limits_{\gamma  \in \prod {\left( {{\bf H}^{\rm A} ,{\bf Z}^{\rm L} } \right)} } \left\langle {D\left( {{\bf D}^{\rm A} ,{\bf D}^{\rm L} } \right) \otimes \gamma ,\gamma } \right\rangle,
	\label{eq:gwd}
\end{equation} 
where $\otimes $ is the Kronecker product, $D\left( { \cdot , \cdot } \right)$ is a distance function defined on the edges of a graph for a $L_2$ distance as:
\begin{equation}
		\left\langle {D\left( {{\bf D}^{\rm A} ,{\bf D}^{\rm L} } \right) \otimes \gamma ,\gamma } \right\rangle  = \sum\limits_{i,j,k,l} {\left| {{ d}_{i,j}^{\rm A}  - {d}_{k,l}^{\rm L} } \right|^2 \gamma _{i,k} \gamma _{j,l} }  \\ 
	\label{eq:dgwd}
\end{equation} 
In Eq. (\ref{eq:dgwd}), the edges of graphs are two distance (or adjacency) matrices defined as (intra-modality distance matrices):
\begin{equation}
	\begin{array}{l}
		{d}_{i,j}^{\rm A}  = {\bf D}^{\rm A} \left( {{\bf h}_i ,{\bf h}_j } \right)\mathop  = \limits^\Delta  1 - \cos \left( {{\bf h}_i ,{\bf h}_j } \right) \\ 
		{d}_{k,l}^{\rm L}   = {\bf D}^{\rm L} \left( {{\bf z}_k ,{\bf z}_l } \right)\mathop  = \limits^\Delta  1 - \cos \left( {{\bf z}_k ,{\bf z}_l } \right) \\ 
	\end{array}
\end{equation}

\subsubsection{Matching on ordered graph: Minimizing Fused WD and GWD}
\label{sect:mfgwd}
Matching on ordered graph is illustrated in Fig. \ref{fig:fig2} (c) where both nodes and edges with ordered topology structure are involved. It is defined as a Fused Gromov-Wasserstein distance (FGWD) \cite{Vayer2020}:
\begin{equation}
	L_{{\rm FGWD}} \mathop  = \limits^\Delta  \alpha L_{{\rm GWD}}  + (1 - \alpha )L_{\rm WD},
\end{equation}
where $\alpha$ is a weighting factor to control the importance between WD and GWD. Substituting the definitions of WD and GWD in Eqs. (\ref{eq:ot}) and (\ref{eq:gwd}), the FGWD is cast to:
\begin{equation}
	L_{{\rm FGWD}} \mathop  = \limits^\Delta  \mathop {\min }\limits_{\mathclap{\gamma  \in \prod {\left( {{\bf H}^{\rm A} ,{\bf Z}^{\rm L} } \right)} }} (1 - \alpha )\left\langle {{\bf D}^{\rm AL},\gamma } \right\rangle  + \alpha \left\langle {D\left( {{\bf D}^{\rm A} ,{\bf D}^{\rm L} } \right) \otimes \gamma ,\gamma } \right\rangle
	\label{eq:LFGWD}
\end{equation}
As the WD problem defined in Eq. (\ref{eq:ot}) can be solved based on fast Sinkhorn algorithm, we can unify the FGWD and approximate it also based on the Sinkhorn algorithm. With reference to the formulation in Eq. (\ref{eq:ot}), we can first estimate an OT coupling with several iterations as \cite{Xie2018}: 
\begin{equation}
	\gamma ^{(t)} \mathop  = \limits^\Delta  \mathop {\arg \min }\limits_{\gamma  \in \prod {\left( {{\bf H}^{\rm A} ,{\bf Z}^{\rm L} } \right)} } \left\langle {{\bf D}_{{\rm FGWD}}^{(t - 1)} ,\gamma } \right\rangle  + \beta {\rm KL(}\gamma |\gamma ^{(t - 1)} {\rm )}
\end{equation}
Based on this $\gamma ^{(t)}$, we can obtain a unified cost as the fused GWD cost as: 
\begin{equation}
	{\bf D}_{{\rm FGWD}}^{(t)} \mathop  = \limits^\Delta  (1 - \alpha ){\bf D}^{\rm AL} + \alpha D\left( {{\bf D}^{\rm A} ,{\bf D}^{\rm L} } \right) \otimes \gamma ^{(t)} 
\end{equation}
Based on this unified cost function, we can directly use the Sinkhorn algorithm as we did in vanilla OT solution in Eq. (\ref{eq:EOTloss}).

\subsubsection{Temporal consistency in graph topology matching}
\label{sect:temporal}

In Section \ref{sect:mfgwd}, graph matching operates on both nodes and edges in acoustic and linguistic spaces while preserving temporal consistency. Since speech sequences have an inherent order, adjacent acoustic frames should align with neighboring linguistic tokens in a sequence. To enforce this, elements with large temporal gaps are less likely to be coupled. Thus, alignment should remain close to the diagonal of temporal coherence. We define the temporal cost matrix ${\bf D}^{{\rm TAL}} $ with elements $d_{i,j}^{{\rm TAL}}  = \left| {\frac{i}{{l_a }} - \frac{j}{{l_t }}} \right|^2$, where $i$ and $j$ index positions in acoustic and linguistic spaces. This temporal constraint is integrated into the OT cost matrix ${\bf \tilde D}^{\rm AL}$ with a control parameter $\rho$ defined as $\tilde d_{i,j}^{{\rm AL}}  = d_{i,j}^{{\rm AL}}  + \rho d_{i,j}^{{\rm TAL}} $,

where $d_{i,j}^{{\rm AL}}$ is an element of cost matrix ${\bf D}^{\rm AL}$ as defined in Eq. (\ref{eq:Cost}). And the Sinkhorn algorithm is applied on the cost function matrix ${\bf \tilde D}^{\rm AL}$ for OT in real implementations.  
\subsection{Loss functions in knowledge transfer learning}
The proposed GM-OT involves two loss functions: the cross-modal alignment and matching loss (right branch of Fig. \ref{fig:fig1}) and the CTC loss (left branch of Fig. \ref{fig:fig1}). In cross-modal alignment, the acoustic feature can be projected onto the linguistic space based on OT coupling as:
\begin{equation}
		{\bf \tilde Z}_{\rm L} \mathop  = \limits^\Delta  {\rm OT}\left( {{\bf H}^{\rm A}  \to {\bf Z}^{\rm L} } \right) = \gamma ^*  \times {\bf H}^{\rm A}  \in \mathbb{R}^{l_t  \times d_t },  
\end{equation}
where $\gamma ^*$ is the OT coupling matrix. Subsequently, the alignment loss is defined as:
\vspace{-2mm}
\begin{equation}
	L_{{\rm align}}  = \sum\limits_{j = 2}^{l_t  - 1} {1 - \cos \left( {{\bf \tilde z}_j^{\rm L} ,{\bf z}_j^{\rm L} } \right)}, 
	\label{eq:Align}
\end{equation}
where ${\bf \tilde z}_j^{\rm L}$ and ${\bf z}_j^{\rm L}$ are row vectors of feature matrices ${\bf \tilde Z}^{\rm L}$ and ${\bf Z}^{\rm L}$, respectively. In Eq. (\ref{eq:Align}), the usage of indices from $2$ to $l_t -1$ is for handling special symbols `$\rm CLS$' and `$\rm SEP$'. For efficient linguistic knowledge transfer to acoustic encoding, the following transforms are designed as indicated in Fig. \ref{fig:fig1}:
\begin{equation}			
		{\bf H}^{{\rm AL}}  = {\bf H}_{{\rm enc}}  + w_s .{\rm LN(FC}_{\rm 3} {\rm (LN(}{\bf H}^{\rm A} {\rm )))} \in \mathbb{R}^{l_a  \times d_a }, 	
	\label{eq:adapter}
\end{equation}
where $w_s$ is a scaling parameter to adjust the importance of transferring linguistic projected feature. Based on this new representation ${\bf H}^{\rm AL}$ which is intended to encode both acoustic and linguistic information, the final probability prediction for ASR is formulated as ${\bf \tilde P} = {\rm Softmax}\left( {{\rm FC1}\left({\bf H}^{\rm AL} \right)} \right)$,

where `FC1' is a linear full-connected transform. The total loss in model training is defined as:
\vspace{-2mm}
\begin{equation}
	L\mathop  = \limits^\Delta  \lambda L_{{\rm CTC}} ({\bf \tilde P},{\bf y}_{{\rm token}} ) + (1 - \lambda ){(L_{{\rm align}}  + L_{{\rm FGWD}} )},   
	\label{eq:totalloss} 	
\end{equation}
where $L_{{\rm CTC}} ({\bf \tilde P},{\bf y}_{{\rm token}} )$ is CTC loss, ${L_{{\rm align}} }$ and ${L_{{\rm FGWD}} }$ are cross-modality alignment loss and GM-OT loss, respectively. After the model is trained, only the left branch of Fig. \ref{fig:fig1} is retained for ASR inference.  

\section{Experiments and results}
\label{sec:exp}
We conducted ASR experiments to evaluate the proposed algorithm on the AISHELL-1 dataset \cite{AISHELL1}, a widely used open-source Mandarin speech corpus. The dataset consists of: A training set with 340 speakers (150 hours of speech). A development set (validation) with 40 speakers (10 hours of speech). A test set with 20 speakers (5 hours of speech). Following the pre-processing strategy in \cite{AISHELL1}, data augmentation techniques were applied to enhance model robustness. Feature settings are configured the same as in \cite{AISHELL1}.
\subsection{Model architecture}
In Fig. \ref{fig:fig1}, the acoustic encoder is formed by stacking $16$ conformer blocks \cite{conformer2020}, with each having a kernel size of 15, attention dimension $d_a=256$, $4$ attention heads, and a 2048-dimensional FFN layer. The `bert-base-chinese' from huggingface is used as the linguistic encoder \cite{Huggingface}. In this Chinese BERT model, $12$ transformer encoders are applied, the token (or vocabulary) size is 21128, and the dimension of linguistic feature representation is $d_t=768$. 
\subsection{Hyper-parameters in model learning}
Several hyper-parameters are associated with the proposed model, and these parameters may have joint effect in efficient linguistic knowledge transfer learning. We empirically set them based on our preliminary experiments. In model optimization, Adam optimizer \cite{Adam} is used with a learning rate (initially set to 0.001) schedule with 20,000 warm-up steps. The model with cross-modal transfer was trained for 130 epochs, and the final model used for evaluation was obtained by averaging models from the last 10 epochs. 
\begin{figure*}[tb]
	\centering
	\includegraphics[width=12cm, height=2cm]{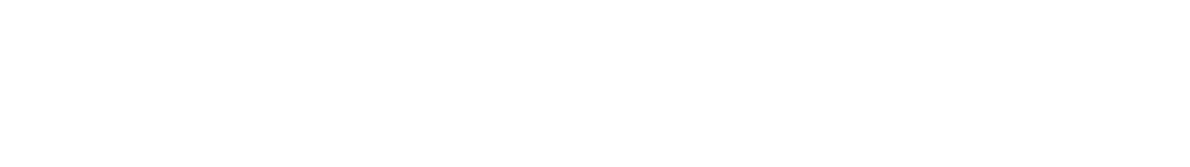}
	\caption{Coupling between acoustic feature and linguistic tokens with changing different parameters, (a) weighting between WD and GWD (varying $\alpha$), (b) controlling temporal consistency between cross-modalities (varying $\rho$) (fixed $\alpha=0.1$), (c) with entropy regularization (varying $\beta$) (fixed $\alpha=0.1$, $\rho=0.5$).}
	\vspace{-4mm}		
	\label{fig:GWD}
\end{figure*}   
\subsection{Behaviors of the GM-OT based feature alignments}
Before conducting ASR experiments, we investigate the behaviors of the GM-OT alignment module with different parameter settings. In this investigation, the acoustic encoder is trained (only with CTC based loss) and fixed for acoustic feature extraction. The coupling between acoustic feature and corresponding linguistic feature sequence is estimated by the GM-OT module. As defined in Eq. (\ref{eq:LFGWD}), when $\alpha=0$, the alignment is reduced to node only. And when $\alpha=1$, the alignment is edge only. With setting different $\alpha$, we can adjust their importance. An example is shown in Fig. \ref{fig:GWD} (a).  

This figure shows the OT couplings between acoustic and its corresponding linguistic tokens of an utterance where horizontal axis is acoustic frame index, and vertical axis denotes linguistic token index. From this figure we can see that increasing weight on edge matching could make even alignment for tokens, i.e., more uniformly segment acoustic features to each token. This is a double-edge sword since not all tokens have uniform lengths of acoustic duration. Furthermore, with changing the parameter $\rho$ ($\alpha=0.1$ is fixed), the effect is showed in Fig. \ref{fig:GWD} (b). From this figure, we can see that emphasizing the temporal coherence along the diagonal direction always constrain the OT coupling to keep monotonic correspondence during alignment.   
Another important parameter is the entropy regularization parameter $\beta$ as used in Eq. (\ref{eq:EOTloss}). An example is showed in Fig. \ref{fig:GWD} (c) with setting different parameter $\beta$ ($\alpha=0.1$, $\rho=0.5$ are fixed). From this figure, we can see that alignment couplings are diffused when increasing the regularization weights. Based on these analysis, we can empirically set parameters in our ASR experiments. 

\subsection{ASR results}
\label{sec:results}
In inference stage, only the left branch (blocks in dashed red box in Fig. \ref{fig:fig1}) is utilized, maintaining the decoding speed similar to that of the CTC-based decoding (CTC greedy search-based decoding was employed in this paper). During experiments, we find that there are several mistakes in the original reference transcriptions for the development and test sets which may affect the fairness of evaluations. Therefore, we checked and corrected the reference transcriptions. All our experimental results are calculated based on the corrected references. Results are showed in table \ref{tab:tab1}. Moreover, we also listed the ASR results of some baseline models Conformer+CTC \cite{Watanabe2017,wenet2.0}, Conformer+CTC/AED \cite{Kim2017, Hori2017,Watanabe2017}, and linguistic knowledge transfer learning-based model NAR-BERT-ASR \cite{NARBERT} (stacking BERT model on acoustic encoder) for comparisons (in light gray). Please be noted that the comparison results are from our own implementations with reference to \cite{Kim2017, Hori2017,Watanabe2017, NARBERT}. And results based on our own implementations are comparable or a little better than the original results. The parameter settings used in table \ref{tab:tab1} are listed in table \ref{tab:tab2}. And in all experiments, the weighting coefficient in Eq. (\ref{eq:totalloss}) is fixed with $\lambda=0.3$. With changes of controlling hyper-parameters, the results of our GM-OT are shown in bottom rows. In this table, as a special case of our GM-OT, Setting 1 is similar as in others work \cite{ASRU2023Lu} which results are comparable. From this table, we can observe that our proposed GM-OT efficiently transfers linguistic knowledge in acoustic encoding, yielding significant performance improvements. 

\begin{table}[tb]
	\centering
	\caption{ASR performance on the AISHELL-1 corpus, CER (\%).}
	\vspace{-2mm}
	\begin{tabular}{|c||c||c|}
		\hline
		Methods &dev set &test set\\	
		\hline		
		\rowcolor{lightgray}
		Conformer+CTC (Baseline)  &5.16 &5.76 \\		
		\hline	
		\rowcolor{lightgray}
		Conformer+CTC/AED (\cite{Watanabe2017,wenet2.0})  &4.31 &4.82 \\						
		\hline
		\rowcolor{lightgray}
		NAR-BERT-ASR (\cite{NARBERT}) &4.18 &4.68 \\
		\hline
		\hline
		GM-OT (Setting 1) (\cite{ASRU2023Lu})  &3.81 &4.19 \\				
		\hline
		GM-OT (Setting 2) &3.70 &4.08 \\
		\hline
		GM-OT (Setting 3) &3.72 &4.11 \\
		\hline
		GM-OT (Setting 4) &\textbf{3.61} &4.04 \\	
		\hline
		GM-OT (Setting 5) &3.65 &\textbf{3.98} \\		
		\hline				
		GM-OT (Setting 6) &3.76 &4.16 \\
		\hline	
		GM-OT (Setting 7)  &3.73 &4.14 \\
		\hline				
		GM-OT (Setting 8)  &3.74 &4.11 \\
		\hline				
	\end{tabular}
	\vspace{-3mm}
	\label{tab:tab1}
\end{table}
\begin{table}[tb]
	\centering
	\caption{Parameter settings in experiments.}
	\vspace{-2mm}
	\begin{tabular}{|c||c||c||c||c|} 
		\hline
		Parameter settings &$\alpha$ & $\rho$ & $\beta$ & $w_s$ \\ 
		\hline
		\hline
		Setting 1 & 0  &0 &0.05 &0.1 \\ 			
		\hline
		Setting 2 &0.01 &0.3 &0.3 &0.05 \\ 
		\hline
		Setting 3 &0.01 &0.5 &0.5 &0.1 \\ 
		\hline
		Setting 4 &0.02 &0.5 &0.5 &0.1 \\ 
		\hline
		Setting 5 &0.02 &0.3 &0.5 &0.1 \\ 
		\hline		
		Setting 6 &0.05 &0.5 &0.5 &0.1 \\ 
	    \hline					
		Setting 7 &0.1 &0.1 &0.3 &0.05 \\ 
		\hline					
		Setting 8 &0.01 &0.5 &0.5 &0.3 \\
		\hline
	\end{tabular}
	\vspace{-5mm}
	\label{tab:tab2}
\end{table}
\vspace{-2mm}
\section{Conclusion and future work}
\label{sec:conclusion}
In this work, we introduced GM-OT, a novel approach that models acoustic and linguistic sequences as ordered graphs for alignment. By structuring feature embeddings as nodes and capturing sequential relationships as edges, GM-OT enables a more comprehensive alignment through graph-based optimal transport. This unified framework subsumes several existing OT-based knowledge transfer methods as special cases. Furthermore, by leveraging transport coupling, GM-OT facilitates the mapping of acoustic features to the linguistic space, allowing for a direct comparison with PLM-encoded information. Our ASR experiments demonstrated the effectiveness of the proposed approach in improving knowledge transfer and overall performance. Despite its advantages, GM-OT introduces several hyper-parameters that are challenging to optimize. Some of these hyper-parameters significantly impact the stability of OT solutions, making model training sensitive to their selections. Robust optimization techniques to enhance the stability and effectiveness of GM-OT remain as our future work.


\bibliographystyle{IEEEtran}

\end{document}